\documentclass[]{article}
\usepackage{amsmath,amssymb,amsbsy,graphicx,setspace,subfigure}
\usepackage[square,numbers,sort&compress,comma,super]{natbib}
\newcommand{\vv}[1]{\ensuremath{\mathbf{#1}}}
\newcommand{\pd}[2]{\ensuremath\frac{\partial #1}{\partial #2}}
\def\rv{\mathbf{r}}

\title{Optimizing directed self-assembled morphology}
\author{Jian Qin, Gurdaman S. Khaira, Yongrui Su, Grant P. Garner\\
\textit{Institute for Molecular Engineering, University of Chicago, Chicago, IL 60637}\\
\\
Marc Miskin, Heinrich M. Jaeger\\
\textit{James Franck Institute, University of Chicago, IL 60637}\\
\\
Juan J. de Pablo\\
\textit{Institute for Molecular Engineering, University of Chicago, Chicago, IL 60637}
}
\date{}

\begin{document}
\maketitle

\begin{abstract}
Directed assembly of block polymers is
rapidly becoming a viable strategy for lithographic patterning of nanoscopic features.
One of the key attributes of directed assembly is that an underlying chemical or
topographic substrate pattern used to direct assembly need not exhibit a direct correspondence with the sought after block polymer morphology,
and past work has largely relied on trial-and-error approaches to design appropriate patterns.
In this work, a computational evolutionary strategy is proposed to solve this optimization problem.
By combining the Cahn-Hilliard equation, which is used to find the equilibrium morphology,
and the covariance-matrix evolutionary strategy,
which is used to optimize the combined outcome of particular substrate-copolymer combinations,
we arrive at an efficient method for design of substrates leading to non-trivial, desirable outcomes.
\end{abstract}

%------------------------------------------------------------------
\section{Introduction}
%------------------------------------------------------------------

Lithography represents one of the key fabrication steps for nanoscopic devices,
ranging from electronic circuits to storage media.\cite{Darling2010}
As critical dimensions continue to shrink,
alternative patterning strategies and materials are being sought
to circumvent some of the patterning challenges that arise at small length scales.
These include roughness, pattern collapse, and defectivity.
In recent years, directed assembly of block copolymers on topographic or chemical patterns
has received considerable attention as a viable and promising patterning approach
for lithographic patterning of ultra-small features.
Block polymers are known to spontaneously self-assemble into a wide range of ordered morphologies,
including lamellar, cylindrical, or spherical structures.\cite{Nealey2007, Darling2010}
In thin films, that self-assembly can be guided through the use of chemical or topographic patterns on the underlying substrate.
Past work has shown that it is possible to direct the assembly of simple diblock copolymers
and their blends with homopolymers into all of the canonical features that arise in integrated circuits,
such as lines, bends, jogs, and spots.
An important concept in directed self-assembly is that of pattern interpolation,
in which only a subset of any desirable features appears on the substrate,
and the block copolymer is used to fill-in the rest, thereby adding information into the fabrication process.
For example, to produce a dense array of parallel lines, one need only use a surface pattern
that includes a fraction of the lines (e.g. one fourth).
Such lines then serve as guiding stripes for a lamellar forming block copolymer
having a period that is four times smaller than the spacing between the surface lines.

%Over the past few years, a variety of nanostructured devices have been fabricated
%with the assistance of block copolymers,
%which were used in magnetic recording,\cite{ChengVancso2001, NaitoKamata2002}
%flash storage,\cite{GuariniGignac2003}
%or capacitor applications.\cite{BlackTuominen2001, BlackMilkove2002}

%From the perspective of applications,
%one challenge of integrating the block copolymer based template technique with
%industry is to obtain a well organized structures over a large area, with low defect density.
%This issue has been addressed by introducing chemical patterns into substrate
%on which the materials are grown,
%or by using solvent annealing.
%These improved processing procedures have enabled the production
%of impressive array of periodically directed self-assembled structures.

%which were used in magnetic recording,
%flash storage,

Dense, periodic arrays of lines or spots are of considerable interest for applications in dense storage media.
\cite{ChengVancso2001, NaitoKamata2002, GuariniGignac2003}
For more complex layouts, such as those encountered in logic devices, a central challenge is to guide the materials to assemble into aperiodic,
more versatile, and more complicated morphologies or geometries. Within the spirit of density interpolation, the underlying pattern used to guide the assembly need not have a one-to-one correspondence with the geometry of interest; the question that arises then is,
for a target morphology, how can one design an optimal sparse pattern
to direct block copolymer self-assembly?

One could of course adopt a traditional inverse Monte Carlo algorithm,
and rely on a random search of suitable patterns to find a plausible solution.
This amounts to a computational trail-and-error search,
akin to that performed in experiments.
Such an approach has been proposed recently,\cite{HannonKatz2013}
with good results, in the context of topographically directed assembly.
In that work, a mean-field theory in two dimensions (SCFT)\cite{Matsen2002rev} was combined with
a random search algorithm to identify topographic features leading to suitable morphologies. The disadvantage of such an approach, however, is that the search is blind and, for large parameter spaces, it can rapidly become intractable.

In this work, we propose to use a state-of-the-art optimization technique, namely
the covariance-matrix adaptation evolutionary strategy (CMA-ES),\cite{EibenSmithBook}
in combination with a mean field copolymer model,
to identify combination optimal chemical patterns for assembly of block polymers into non-regular morphologies.

%Cahn-Hilliard equation augmented with
%the Ginzburg Landau free energy that is tailored to block copolymer,

%The Cahn-Hilliard equation is used to explore the block copolymer self-asembling morphology,
%and the CMA-ES is used to optimize patterns.
%The details of our methodology and results are explained in secs. \ref{sec:method}
%and \ref{sec:result} respectively.
%In sec. \ref{sec:result}, the performance of our methods is demonstrated
%by using three nontrivial examples.
%Sec. \ref{sec:summary} is the summary and outlook.

%------------------------------------------------------------------
\section{Methodology}
\label{sec:method}
%------------------------------------------------------------------

For concreteness, the strategy presented in this work is described
in the context of a surface pattern consisting of circular spots. Extensions to other types of patterns are trivial.
The goal is to use the minimal number of spots on the surface to
direct the assembly of a lamellar forming diblock copolymer into a target morphology.
To determine the equilibrium copolymer morphology for a given placement of the surface spots,
we use Ginzburg-Landau (GL) free energy functional,
and we evolve morphology using the Cahn-Hilliard (CH) equation.\cite{CahnHilliard1958}
One could use other approaches, including more elaborate mean field theories,
theoretically-informed approaches, or full-blown molecular simulations,
but a generic GL-CH approach is simple and allows us to demonstrate
the general principles put forth in this work without loss of generality.
Furthermore, it facilitates studies of large systems,
and enables simulations of the evolution dynamics of any given morphology.
\cite{QiWang1999, YamadaTakao2004}

In order to evolve a particular combination of spots and morphology in parameter space, we
introduce an objective or ``fitness'' function that quantifies the difference between the equilibrium,
instantaneous morphology and the target morphology. That function depends on the spot positions,
with the number of spots held constant. The fitness function is then
minimized with a covariance matrix adaptation evolutionary strategy (CMA-ES).\cite{EibenSmithBook}
CMA-ES is based on the idea of ``natural evolution'', and has proven to be particularly
efficient for optimization of complicated functions when little is known about the underlying landscape.\cite{EibenSmithBook}

%------------------------------------------------------------------
\subsection{Cahn-Hilliard equation for assembled morphology}
\label{subsec:chequation}
%------------------------------------------------------------------

For simplicity, we consider a system of pure diblock copolymers composed of A and B blocks.
A and B type monomers have the same reference volume and statistical segment length.
The total number of beads and the volume fraction of A blocks are denoted $N$ and $f$, respectively.
The excess free energy cost of creating an A/B contact is quantified by the Flory-Huggins parameter $\chi$.
The product $\chi N$ controls the degree of phase separation;
the higher its value, the stronger the tendency of A and B blocks are to segregate.

Following Shi et al.,\cite{LiShi2010, XieShi2013}
we use the free energy form developed by Ohta and Kawasaki \cite{OhtaKawasaki1986}
to characterize the system morphology. This formalism
is valid in the strong segregation regime (large $\chi N$);
it is expressed as the sum of three terms
\begin{equation}
   F[\phi] = F_\text{GL}[\phi] + F_\text{non-local}[\phi] + \int d\rv H_\text{ext}(\rv) \phi(\rv).
   \label{eq:freeenergy}
\end{equation}
Here $\phi(\rv)$ is the order parameter field quantifying the extent of phase separation,
defined as the monomer volume fraction difference, $\phi(\rv) \equiv \phi_A(\rv) - \phi_B(\rv)$.
$H_\text{ext}(\rv)$ is the external potential representing the interaction
between the guiding spots and the copolymer; we use the hyperbolic tangent function introduced in ref. \cite{LiShi2010},
$H_\text{ext}(\rv) = -(1/2) V_0 (\tanh (-|\rv - \vv{R}| + \sigma)/\lambda + 1)$,
where $\vv{R}$ is the position of the spot center, $V_0$ and $\sigma$ are the strength
and range of the potential, and $\lambda$ controls the steepness of the potential's decay.

\begin{figure}[htb]
\centering
\includegraphics[width=.45\textwidth,height=!]{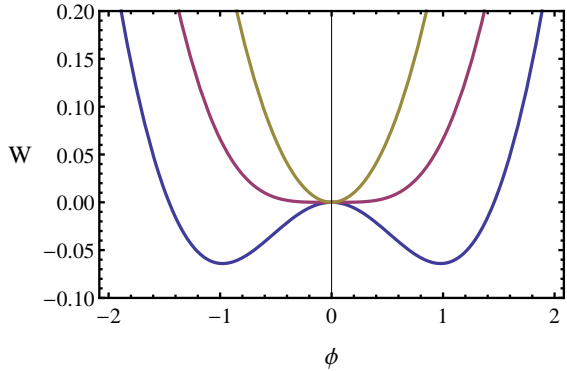}
\caption{Local demixing free energy as a function of order parameter.
Yellow: $A=0.5$; red: $A=1.0$; blue: $A=1.3$.
}
\label{fig:WPhi}
\end{figure}

The Ginzburg-Landau free energy $F_\text{GL}[\phi]$ can be written as
\begin{equation}
   F_\text{GL}[\phi] = \int d\rv \left[ \frac{1}{2} \left(\nabla\phi\right)^2 + W(\phi) \right].
   \label{eq:GLfreeenergy}
\end{equation}
The gradient term represents the free energy cost associated with spatial inhomogeneities.
The $W(\phi)$ term is the local free energy density that drives the phase separation.
It depends on the Flory-Huggins $\chi$-parameter,
and contains only even powers of $\phi$
(this is true since exchanging A and B block labels has no physical consequence).
$W$ is generally assumed to be of the form $(1/2)\phi^2 + (g/4)\phi^4$.
In this work, following Shi et al. \cite{LiShi2010},
we set $W = -A \ln \cosh(\phi) + \phi^2/2$.
Here the parameter $A$ controls the degree of phase separation.
$W$ has one minimum at $\phi = 0$ for $A < 1$ and has two minima for $A > 1$.
The shapes of $W$ at $A = 0.5$, 1.0, and 1.3 are shown in Fig. \ref{fig:WPhi}.

The term $F_\text{non-local}$ is the chain stretching energy describing the chain connectivity.
It can be written as \cite{OhtaKawasaki1986, LiShi2010}
\begin{equation}
   F_\text{non-local} = \frac{\alpha}{2} \int d\rv \int d\rv' \delta\phi(\rv) G(\rv-\rv') \delta\phi(\rv'),
   \label{eq:nonlocal}
\end{equation}
where $\delta\phi(\rv) \equiv \phi(\rv) - \bar\phi$ is the deviation from the homogeneous value, and $\bar\phi = 2f - 1$.
$G(\rv, \rv')$ satisfies $-\nabla^2G(\rv,\rv') = \delta(\rv-\rv')$.
For simplicity, we consider a two-dimensional representation here, where $G(\rv, \rv') = -\ln(|\rv - \rv'|)/2\pi$.\cite{OhtaKawasaki1986}

Equations (\ref{eq:GLfreeenergy}) and (\ref{eq:nonlocal}) are purely phenomenological.
The mean field free energy for a diblock copolymer melt may be mapped onto this form
by using the explicit expressions in Ref. \cite{OhtaKawasaki1986} (Eqs. (4.5-7)).
By inspection, one can arrive at the following mapping rules:
(1) the length scales in Eqs. (\ref{eq:GLfreeenergy}) and (\ref{eq:nonlocal}) are in units of $\xi_0$,
where $\xi_0^2$ is defined by $R_g^2/(4 f (1-f) \chi_sN)$ and
$R_g$ is the radius of gyration of the diblock copolymer;
(2) the value of $A$ equals $\chi/\chi_s$,
where $\chi_s$ is the value of the mean field spinodal ($\chi_s = 10.5/N$ for $f=0$);
(3) the value of $\alpha$ equals $3/(16 f^3 (1-f)^3 (\chi_sN)^2)$.
The extent of the phase separation is controlled by the value of $\chi$ or $A$.
The equilibrium domain spacing is proportional to $N^{2/3}(A-1)^{1/6}$.\cite{OhtaKawasaki1986}

To find the equilibrium morphology, we use the Cahn-Hilliard equation to evolve the $\phi$ field,
which is appropriate for conserved order parameters
and has been widely used to study the material structural evolution in phase field models.\cite{Chen2002}
The Cahn-Hilliard equation has the form
\begin{equation}
   \pd{\phi(\rv,t)}{t} = M \nabla^2 \frac{\delta F[\phi]}{\delta \phi(\rv,t)}, %+ \eta(\rv,t),
\end{equation}
where $M$ is the effective mobility coefficient and is set to unity.
Substituting the free energy expression into the Cahn-Hilliard equation, we get\cite{LiShi2010}
\begin{align}
   \pd{\phi(\rv,t)}{t} = &
    \nabla^2 \left(- \nabla^2 \phi(\rv, t) - A \tanh(\phi(\rv,t)) + \phi(\rv,t) \right) \nonumber\\
   & + \nabla^2 H_\text{ext}(\rv)
   - \alpha \delta\phi(\rv,t).
   \label{eq:CHequation}
\end{align}
%Here $\eta$ is the noise term and satisfies
%$\ave{\eta(\rv,t)\eta(\rv',t')} = - \eta_0 \nabla^2 \delta(\rv-\rv') \delta(t-t')$.
Since the Cahn-Hilliard equation is essentially a diffusion equation,
the total monomer content in the system remains constant as time evolves, i.e.,
$\int d\rv \phi(\rv,t) = \bar\phi V$, where $V$ is the volume of the system.
For a given block volume fraction $f$ and given spot positions,
the values of $A$, $\alpha$, and $H_\text{ext}(\rv)$ are fixed. The
effect of $f$ is implicit in the $\delta\phi$ term.
For random initial field values satisfying the stoichiometrical constraint,
after a sufficiently long time,
the Cahn-Hilliard equation will typically evolve the system into a local equilibrium state.

%------------------------------------------------------------------
\subsection{CMA-ES optimization}
%------------------------------------------------------------------

Let the target morphology be described by $\tilde\phi(\rv)$, and
the equilibrium morphology under a given set of spot constraints
be $\phi(\rv; \{\vv{R}_i\})$, where the dependence on pole position vectors $\{\vv{R}_i\}$ is explicitly shown.
The difference between $\phi$ and $\tilde\phi$ can be quantified by
\begin{equation}
   \Omega(\{\vv{R}_i\}) \equiv \int d\rv \left(\phi(\rv; {\vv{R}_i}) - \tilde\phi(\rv) \right)^2.
   \label{eq:objective}
\end{equation}
Our goal is to optimize spot positions by minimizing $\Omega$.

We resort to the covariance matrix adaptation evolutionary strategy or CMA-ES to minimize $\Omega$.
CMA-ES belongs to a family of evolutionary optimization algorithms that
mimic the principle of biological evolution.\cite{EibenSmithBook}
Recently, it
has been used with considerable success in the context of materials research
for optimization of packing problems\cite{MiskinJaeger2013}
and for crystal structure prediction\cite{OganovValle2001}
(the use of different variants of evolutionary algorithm have
also been reported \cite{BianchiKahl2012, DennisonDijkstra2012}).
It is iterative, stochastic, and does not require that the derivative of the objective function be evaluated.
At each iteration stage or generation,
a finite number ($\lambda$) of samples derived from the previous generation is allowed
to mutate and recombine following a prescribed protocol;
these offspring are then ranked according to the objective function, and
the ``best'' $\mu$ offspring are used for the next generation iteration.

The key to implementing such an algorithm is designing an efficient protocol for mutation and recombination,
which on the one hand maintains the population diversity, so that the system is not trapped into local extrema,
and on the other hand ensures fast convergence in the neighborhood of the optimal extremum.
In a naive random search, the older population is perturbed by independently distributed Gaussian random numbers.
In CMA-ES, the correlation among different searching directions, as measured by the covariance matrix,
is explicitly considered,
and the covariance matrix adapted at each iteration step by ``learning'' from the fitness of the entire population.
The idea is analogous to the approximation of the Hessian matrix in the quasi-Newton method
in deterministic optimization.

\begin{figure}[htb]
\centering
\includegraphics[width=.45\textwidth,height=!]{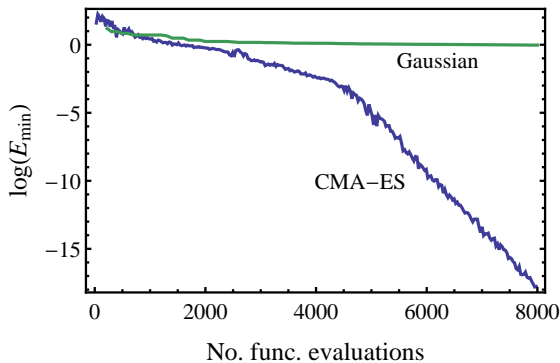}
\caption{Convergence behavior of CMA-ES compared to a Gaussian random search,
tested on the 12-dimensional Rosenbrock function,
which has a global minimum at zero when all its arguments equal unity.
The range for the Gaussian random search is varied using the ``1/5'' acceptance rate rule \cite{EibenSmithBook}.
The CMA-ES uses a evolution population of 28 members, out of which 4 best members are selected at each generation.
In both cases, the initial point is generated at random.
}
\label{fig:CMAESGaussian}
\end{figure}

Our implementation of the CMA-ES is based on the improved algorithm discussed in Ref. \cite{Hansen2003}.
Most parameters required by the algorithm have been set on the basis of heuristic arguments.
For our problem, the population size $\lambda$ and
the number of offspring used to generate new populations $\mu$ are 28 and 4, respectively
(the ratio of the two was recommended to be 7 \cite{Hansen2003}).
To examine the convergence behavior of CMA-ES, in Fig. \ref{fig:CMAESGaussian},
we compared the optimization results obtained from CMA-ES and the Gaussian random search
for the 12-dimensional Rosenbrock function.
Although the exact shape of the convergence curve depends slightly on the location of the starting point,
the curves in the figure are representative of the typical efficiency for both methods.
It is apparent that the Gaussian random search is frequently trapped
in local minima,
whereas the CMA-ES is able to find the global minimum. Furthermore,
the convergence rate is exponential after a sufficient number iterations.

%------------------------------------------------------------------
\section{Results}
\label{sec:result}
%------------------------------------------------------------------

For the problem of interest here, solving the CH equation is the most computationally demanding step.
With that issue in mind, in Sec. \ref{subsec:chalgorithm} we focus on the algorithm and optimize the parameters
used to solve the CH equation.
Then, in Sec. \ref{subsec:optimization},
we present the optimization results obtained using the CMA-ES algorithm.

%------------------------------------------------------------------
\subsection{Evolving Cahn-Hilliard equation}
\label{subsec:chalgorithm}
%------------------------------------------------------------------

\begin{figure}[htb]
\centering
\includegraphics[width=.45\textwidth,height=!]{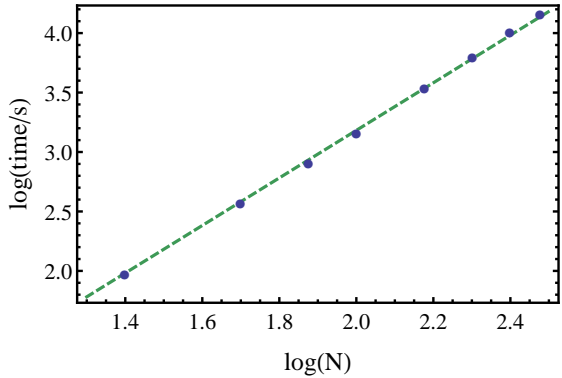}
\caption{
Numerical complexity of the algorithm for the CH equation.
The abscissa correspond to the number of grid point per edge $N$.
The ordinate axis corresponds to time spent on $2\times10^5$ iteration steps.
The dashed line has a slope of 2.
Parameters: $A = 1.3$, $\alpha=0.002$, $f=0.5$, and $\delta t=0.02$.
}
\label{fig:CHtiming}
\end{figure}

To solve the CH equation, we discretize the square-shaped simulation cell into an $N$-by-$N$ grid,
and approximate the gradient term in Eq. (\ref{eq:CHequation}) using central differences.
The composition field $\phi(\rv,t)$ is propagated in time using the forward Euler's method,
with $\delta t$ as the time step (effectively, the mobility factor $M$ can be absorbed into
the definition of $\delta t$).
The algorithm complexity scales with $N^2$, as confirmed by Fig. \ref{fig:CHtiming},
in which the time spent on a fixed number of CH equation iterations is plotted versus
the number of grid points $N$, on a logarithmic scale.
The results can be fit with a straight line of slope 2.

\begin{figure}[htb]
\centering
\includegraphics[width=.45\textwidth,height=!]{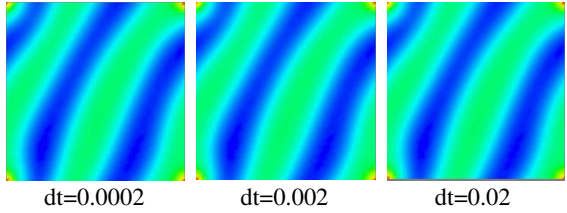}
\caption{Effects of $\delta t$ on equilibrium morphology.
Equilibrium morphologies obtained by propagating the CH equation
with three different values of time step: 0.0002, 0.002, and 0.02.
Parameters: $A=1.3$, $\alpha=0.02$, $N=50$.
}
\label{fig:dtConvergence}
\end{figure}

To find a proper value of time step $\delta t$ that is
small enough to ensure numerical stability
yet large enough to evolve the CH equation efficiently,
we compared results attained at different $\delta t$ values.
Fig. \ref{fig:dtConvergence} shows equilibrium morphologies obtained
by solving the CH equation using three different values of $\delta t$,
starting from the same initial configuration.
The fact that the three morphologies are nearly indistinguishable
suggests that using $\delta t = 0.02$ is sufficient.
In the remainder of this work, we use this value for our calculations.

\begin{figure}[htb]
\centering
\includegraphics[width=.45\textwidth,height=!]{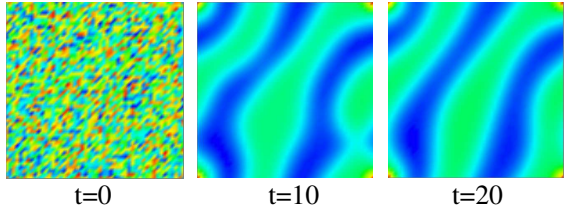}
\caption{Time dependence of the morphology evolution.
Morphologies at different times along the same evolution trajectory using $\delta t = 0.02$.
Parameters: $A = 1.3$, $\alpha = 0.002$, $N = 50$.
}
\label{fig:timeSequence}
\end{figure}

The other parameter to be optimized is the number of iteration steps.
In our study, we want this number to be sufficiently large to allow the system to
reach the equilibrium morphology for a given arrangement of the spots;
however, we also want to avoid spending time on equilibrated morphologies.
Fig. \ref{fig:timeSequence} shows the morphologies at different times along the same trajectory.
It is clear that after at least 1000 iteration steps, which correspond to $t=20$,
the equilibrium morphology has been found.
For the results presented in the next section, we used an iteration number of 4000,
which ensures that local equilibrium morphologies are found.

%------------------------------------------------------------------
\subsection{Phase diagram in $A-f$ plane}
%------------------------------------------------------------------

\begin{figure}[htb]
\centering
\includegraphics[width=.45\textwidth,height=!]{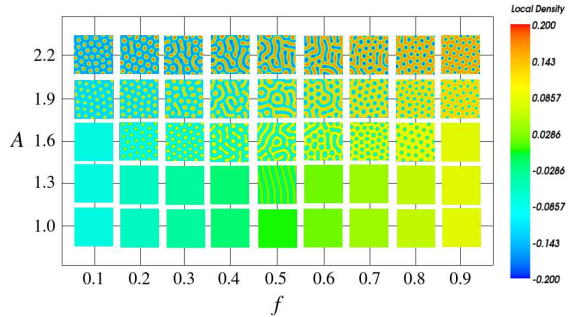}
\caption{2-D phase diagram at varying values of $A$ and $f$, for $\alpha=0.02$.
Morphologies are obtained by evolving the CH equation for $2\times10^8$ steps
from a random initial configuration ($\delta t = 0.01$).
The grid size: $100\times100$.
}
\label{fig:CHphase}
\end{figure}

Before optimizing the spot positions using the evolutionary algorithm,
we explored the effects of various controlling parameters in the generic CH equation.
Fig. \ref{fig:CHphase} shows the typical morphologies obtained
for various values of $A$ and $f$, and at a fixed value of $\alpha$.
As discussed in Sec. \ref{sec:method},
$f$ is the block volume fraction that controls the symmetry of the morphology,
and $A$ is analogous to the $\chi$ parameter, which controls the strength of block incompatibility.

The results in Fig. \ref{fig:CHphase} are consistent with the physical meaning of $A$ and $\alpha$.
For $A \leq 1.0$, homogeneous morphologies are found for all $f$ values.
For $A > 1.0$, the lamellar patterns are found at compositions close to $f=0.5$,
and the hexagonally packed cylindrical patterns are found at asymmetric compositions,
even though in both cases the presence of defects is apparent.

In what follows, we focus on systems forming a lamellar morphology,
and used the following parameters: $A=1.3$, $f=0.5$, $\alpha=0.002$.
This set of parameter gives distinct lamellae having a natural periodicity of $L_0 \simeq 20$.

%------------------------------------------------------------------
\subsection{Optimization using evolutionary algorithm}
\label{subsec:optimization}
%------------------------------------------------------------------

\begin{figure}[htb]
\centering
\includegraphics[width=.45\textwidth,height=!]{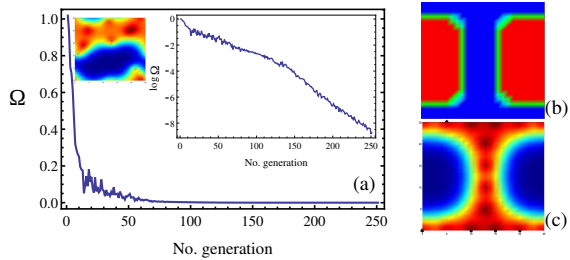}
\caption{Evolutionary results of ``I'' pattern.
(a) Evolution of the objective function;
(b) The target morphology.
(c) The optimal morphology and the spot positions.
Parameters: $A=1.3$, $\alpha=0.002$, $N=50$.
}
\label{fig:Ievolution}
\end{figure}

We now present results obtained using the CMA-ES optimization.
As mentioned above, the population size is $\lambda = 28$,
and the number of fitting samples used to spawn new trajectories is $\mu = 4$.
The first target morphology defined here is a pattern mimicking the letter ``I'',
shown in Fig. \ref{fig:Ievolution}(b).
The number of anchoring spots representing the chemical pattern is 9.
Initial spot positions are generated at random, and the initial morphology
is calculated by evolving the CH equation, as shown in the inset of \ref{fig:Ievolution}(a).
At each iteration step, the spots are repositioned using the CMA-ES algorithm,
and the equilibrium morphology is generated by solving the CH equation.
The values of the objective function are calculated using Eq. (\ref{eq:objective}) and are
plotted in Fig. \ref{fig:Ievolution}(a), as a function of iteration number
(also see the inset for a plot on a logarithmic scale).
The results suggest that the magnitude of the objective function decays nearly exponentially,
and that there exist two convergence rate regimes.
The first (below 150 iterations) has a smaller slope;
the second regime (above 150 iterations) has a greater slope.
The existence of these two regimes mimics the behavior shown in Fig. \ref{fig:CMAESGaussian},
implying that the spot positions are first optimized globally, and then locally.
The results also show that the optimal spot positions are identified within 250 evolution iterations,
and that the residual value of the objective function drops to the level of $10^{-8}$.
The final configuration and the corresponding spot positions are shown in Fig. \ref{fig:Ievolution}(c).

%\begin{figure}[htb]
%\centering
%\includegraphics[width=.4\textwidth,height=!]{predefinedPattern}
%\caption{Morphology evolution results from a predfined pattern.
%}
%\label{fig:predefinedpattern}
%\end{figure}

To verify that the solution identified by CMA-ES is indeed at least a local optimal,
we performed the following test: we
first place the spots at ideal positions that are likely to generate the ``I'' pattern,
and then use the solution of the CH equation as the target morphology and re-iterate from random state.
Since now the target morphology is a solution of the CH equation,
it is also a well-defined minimum of the objective function Eq. (\ref{eq:objective}),
and ideally the minimum should be bracketed by the CMA-ES algorithm.
This is indeed confirmed by our results.
On the other hand, in general, the exact spot positions obtained from CMA-ES optimization
depend slightly on the initial configuration.
One way to reduce this dependence is to conduct multiple optimizations,
and use the average.

\begin{figure}[htb]
\centering
\includegraphics[width=.45\textwidth,height=!]{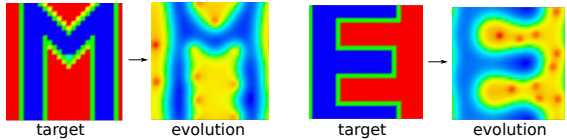}
\caption{Target and optimal morphologies for the ``M'' and ``E'' patterns.
The parameter set is the same as Fig. \ref{fig:Ievolution}.
}
\label{fig:MEevolution}
\end{figure}

To further test the efficiency of the CMA-ES algorithm, we used several other nontrivial patterns.
Two sets of target and optimized morphologies are shown in Fig. \ref{fig:MEevolution}, which
mimic the letters ``M'' and ``E'', respectively.
Both of these two patterns are generated using the same parameter set as the pattern ``I'',
and the convergence behaviors are similar.

%------------------------------------------------------------------
\section{Summary}
\label{sec:summary}
%------------------------------------------------------------------

We have presented a methodology to solve the pattern design problem by
using a Cahn-Hilliard equation to find the equilibrium morphology of diblock copolymers and
by using the CMA-ES algorithm to optimize the underlying chemical pattern.
The applicability and usefulness of the proposed strategy were demonstrated for lamellar forming diblock copolymers,
and three nontrivial target morphologies.

The size of the systems considered here was modest, about $2L_0 \times 2L_0$,
and the overall calculation time required to generate an optimal solution was approximately 8 hours on a single processor.
Extension of the methodology to larger systems and different morphologies is straightforward.
the computational efficiency of the proposed approach could be easily increased by using parallel algorithms:
(1) As shown in Fig. \ref{fig:CHtiming}, the numerical complexity for solving
the CH equation scales with the system size ($N^2$). This step involves essentially
matrix-vector products, and can be readily parallelized.
(2) The CMA-ES essentially involves a set of independent populations,
which can also be parallelized in a trivial way.

The objective function used in this work is the simplest that one can think of.
More elaborate versions could of course be used.
For instance, instead of calculating the difference in real space,
one may consider the difference in Fourier mode coefficients.
Assigning different weights to long and short wavelength modes
may lead to more efficient optimization behavior.

The Cahn-Hilliard equation was used in this work to resolve the composition profile.
As a generic framework, the equation also enables us to study assembly dynamics,
and can be adapted to study more complex systems, including polymer blends.\cite{XieShi2013}
These possibilities will be addressed in future work.

\textbf{Acknowledgement}

%-----------------------------------------------------------
% bibliography
%-----------------------------------------------------------
\renewcommand\refname{Notes and references}
\bibliographystyle{rsc}
%\bibliography{ref/esch.bib}
\bibliography{esch.bib}

\end{document}